\documentclass[12pt
               twocolumn]{iopart}
\usepackage{graphicx}
\usepackage{appendix}
\newcommand{\be}{\begin{equation}}
\newcommand{\ee}{\end{equation}}
\newcommand{\ba}{\begin{eqnarray}}
\newcommand{\ea}{\end{eqnarray}}

\usepackage{color,ulem}

\begin{document}

\title[Testing general relativity with compact coalescing binaries:
  exact vs predictive methods]{Testing general relativity with compact coalescing binaries:
  comparing exact and predictive methods to compute the Bayes factor}
\author{Walter Del Pozzo \footnote{{wdp@star.sr.bham.ac.uk}}, Katherine Grover, Ilya Mandel, Alberto Vecchio}
\address{School of Physics and Astronomy, University of Birmingham, Edgbaston, Birmingham B15 2TT, United Kingdom}

\begin{abstract}
The second generation of gravitational-wave detectors is scheduled
to start operations in 2015.  Gravitational-wave signatures of compact binary coalescences could be used to accurately test the strong-field dynamical predictions of general relativity.  
Computationally expensive data analysis pipelines, including TIGER, have been developed to carry out such tests.   As a means to cheaply assess whether a particular
deviation from general relativity can be detected, Cornish et al.~\cite{cornish:2011} and Vallisneri \cite{vallisneri:2012}
recently proposed an approximate scheme to compute the Bayes factor
between a general-relativity gravitational-wave model and a model representing a
class of alternative theories of gravity parametrised by one
additional parameter.  This approximate scheme is based on only two easy-to-compute quantities: the signal-to-noise ratio of the signal and the fitting factor between the signal and the manifold of possible waveforms within general relativity.

In this work, we compare the prediction from the approximate formula against
an exact numerical calculation of the Bayes factor using the
\texttt{lalinference} library.  We find that, using frequency-domain waveforms, 
the approximate scheme predicts 
exact results with good accuracy, providing the correct scaling with the signal-to-noise ratio at a fitting factor value of $0.992$ and the correct scaling with the fitting factor at a signal-to-noise ratio of $20$, down to a fitting factor of $\sim 0.9$.  We extend the framework for the approximate calculation of the Bayes factor which significantly increases its range of validity, at least to fitting factors of $\sim 0.7$ or higher.
\end{abstract}

\maketitle

\section{Introduction}

The upgraded versions of the ground-based gravitational
wave detectors LIGO \cite{ILigo,ALigo} and Virgo \cite{IVirgo,IVirgo2,IVirgo3,AVirgo} are expected to detect gravitational-wave signals from the coalescence of compact binary systems. The prospect of frequent detections, with expected rates between one per few years and a few hundred per year \cite{rates}, promises to yield a variety of
scientific discoveries. 
Among these, the possibility of testing the strong field dynamics of
general relativity (GR) has received increasing attention (e.g., \cite{delpozzo:2011,cornish:2011,sampson:2013,li:2012a,li:2012b,agathos:2013}). 
In fact, during the latest phase of the inspiral, typical orbital velocities are an
appreciable fraction of the speed of light ($v/c\sim 0.4$); following merger, the 
{\it compactness} $G M/(R c^2)$ of the newly formed black hole that is undergoing quasinormal ringing is close to 1.   By
comparison, the orbital velocity of the double pulsar J0737-3039 
is $O(10^{-3}c)$ and its compactness is $\sim 10^{-6}$
\cite{burgay:2003}.
Consequently, efforts have concentrated on the development of robust frameworks
to reliably detect deviations from GR using gravitational-wave signatures of compact-binary mergers.

One of these frameworks is the so-called {\it Test
  Infrastructure for General Relativity} (TIGER)
\cite{li:2012a,li:2012b,agathos:2013}. 
TIGER operates by computing the odds ratio between GR
and a test model in which one or more of the post-Newtonian coefficients are
allowed to deviate from the value predicted by GR. The interested reader is
referred to Refs.~\cite{li:2012a,li:2012b,agathos:2013} for the
details of the method and for analysis of its robustness against various potential systematic
effects. 
To account for unmodelled effects, TIGER constructs a
``background'' distribution of odds ratios between GR and the test
hypothesis by analysing $O(10^3)$ simulated GR signals.
The background distribution defines the null hypothesis against which
any particular observation (or catalog of observations) is tested. 
For validation purposes, the sensitivity of the algorithm to a specific
deviation from GR is currently assessed by comparing it with a
``foreground'' odds ratio distribution. The foreground distribution is
constructed by simulating a variety of signals in which the chosen
deviation from GR is introduced. If the {\it integrated} overlap
between the foreground and background distributions is smaller than a given false alarm
probability, sensitivity to that particular deviation can be claimed.

The process described in the
previous paragraph is extremely computationally expensive.  If arbitrary combinations of $k$
post-Newtonian coefficients are allowed to deviate from GR values, the total number of simulations that necessary to construct the background is $2^k$ for each synthetic
source.  

As a means to cheaply evaluate the detectability of particular
deviations from GR, Cornish et al.~\cite{cornish:2011} proposed an
approximate formula to calculate the odds ratio between GR and an
alternative model for gravity (AG). Subsequently, Vallisneri
\cite{vallisneri:2012} proposed a similar approximation derived from the Fisher
matrix formalism. Vallisneri's
approximation considers the distribution of the odds ratio in the presence of noise
and caracterises the efficiency and false alarm
of a Bayesian detection scheme for alternative theories of
gravity.  Whilst neither of these approaches can replace the necessary
analysis for real data, the possibility of
having a quick and easily understandable formalism to check the
performance of complex pipelines such as TIGER and assess whether a
specific type of deviation is detectable \emph{without} having to
run thousands of simulations seems quite attractive. 

In this work we investigate, in an idealised and controlled scenario,
whether the predictions from Refs.~\cite{cornish:2011,vallisneri:2012} are in agreement with
the output of a numerical Bayesian odds-ratio calculation. We find in particular that the
the analytical prescription of Ref.~\cite{vallisneri:2012} is in reasonable agreement with the
numerical result when the fitting factor (FF) between AG and GR
waveforms is $\geq 0.9$, and that for FF $\leq 0.8$, both analytical
prescriptions overestimate the exact odds ratio. 

Nevertheless, when the analytical odds ratio is regarded as an upper
limit, useful indications of the detectability of a given deviation
from GR can be drawn. 

We analytically correct the approximate framework for computing the
Bayes factor by introducing terms that are negligible at $FF\sim 1$,
reproducing the proposed analytical expressions given in
\cite{cornish:2011,vallisneri:2012}, but become significant at lower
values of the fitting factor.  
We show that these corrections extend the range of validity of the approximate expressions at least down to fitting factor values of $\sim 0.7$.

The rest of the paper is organised as follows: in section
\ref{sec:bayes} we briefly review the Bayesian definition of the odds
ratio; in section \ref{sec:anal} we introduce the formula from
Ref.~\cite{vallisneri:2012}. In section \ref{sec:results} we present
our findings and finally we discuss them in section \ref{sec:discussion}.
	
\section{Bayesian Inference for gravitational wave signals} \label{sec:bayes}

In a Bayesian context, the relative probability of two or more
alternative hypotheses given observed data $d$ is described by the
odds ratio (see, e.g., \cite{delpozzo:2011}).  If GR is the general relativity hypothesis and AG is the hypothesis corresponding to some alternative theory of gravity, the odds ratio is given by:
	\be
		O_{\mathrm{AG},\mathrm{GR}} = \frac{ p(\mathrm{AG} | d) } { p(\mathrm{GR} | d) } =
                \frac{ p(\mathrm{AG}) } { p(\mathrm{GR}) } \frac{ p(d | \mathrm{AG})}{ p(d|\mathrm{GR})}\equiv \frac{ p(\mathrm{AG}) } {
                  p(\mathrm{GR}) } B_{\mathrm{AG},\mathrm{GR}}
		\label{odds}
	\ee
where we introduced the Bayes factor $B_{AG,GR}$, which is the ratio of the
marginalised likelihoods (or evidences). The marginal likelihood is the expectation value  of
the likelihood of observing the data given
the specific model $H$ under consideration over
of the prior probability distribution for all
the model parameters $\theta$:
\be\label{eq:marg-likelihood}
p(d | H) \equiv Z =\int d\theta\, p(d|\theta,H)\, p(\theta|H)\,.
\ee
With the exception of a few idealised cases, the integral
(\ref{eq:marg-likelihood}) is, in general, not tractable
analytically. In gravitational-wave data analysis, the parameter
space is at least 9-dimensional (for binaries with components that are assumed to have zero spin),
and up to 15-dimensional for binaries with arbitrary component spins, and the integrand is a complex
function of the data and the waveform model. For stationary Gaussian noise
\be
p(d|\theta,H) \propto \exp[-(d-h(\theta)|d-h(\theta))/2]\, ,
\ee
where $h(\theta) \equiv h(\theta|H)$ is the model waveform given parameters $\theta$ and we introduced the scalar product
\be
(a|b)\equiv 2\int_0^{\infty} df\, \frac{a(f)b(f)^*+a(f)^* b(f)}{S(f)}
\ee
with the one-sided noise power spectral density $S(f)$. We analysed
data from a single detector with a noise spectral density corresponding to the zero-detuning,
high-power Advanced LIGO design configuration \cite{PSD:AL}.

	\subsection{Analytical Approximation}\label{sec:anal}

Vallisneri \cite{vallisneri:2012} proposed an analytical approximation
to the integral (\ref{eq:marg-likelihood}). He considered the
following assumptions:
\begin{itemize}
\item linear signal approximation leading to a quadratic approximation of the log likelihood;
\item only one additional dimension is necessary to describe the AG model;
\item uniform prior distributions for all parameters describing both GR and AG
  models;
\item the distance between the AG waveform and the manifold of
  GR waveforms is small so that the fitting factor (FF) between the two, defined as
	\be\label{eq:ff}
        FF=\left[ \frac{(h_{\mathrm{AG}}|h_{\mathrm{GR}}(\theta))}{\sqrt{(h_{\mathrm{AG}}|h_{\mathrm{AG}}) (h_{\mathrm{GR}}(\theta)|h_{\mathrm{GR}}(\theta))}}\right]_{\mathrm{max\,over\,} \theta}\,,
	\ee
is close to unity.  
\end{itemize} 
With the above assumptions, the integral Eq.~(\ref{eq:marg-likelihood})
can be approximately computed analytically and the Bayes factor (\ref{odds}) is then given
by: 
	\be
		B_{\mathrm{AG},\mathrm{GR}} \approx \sqrt{2 \pi}  \frac{ \Delta \theta ^a _{est}} {\Delta \theta^a _{prior}} e^{\rho^2 (1 - FF)} \,,
		\label{odds approx}	
	\ee
where $\rho$ denotes the optimal signal-to-noise ratio
(SNR)\footnote{Note that the definition by Vallisneri of the
  signal-to-noise ratio is different from ours. In
  Ref.~\cite{vallisneri:2012} the signal-to-noise ratio quantity that
  appears in Eq.~(\ref{odds approx}) is the norm of an hypothetical GR
signal whose parameters are exactly the same as the ``true'' AG
waveform, but with the extra AG parameter set to zero. In our case,
the signal-to-noise ratio corresponds to the power in the AG signal (in the ideal case when it is filtered with AG templates). However, when
the AG parameter is present only in the phase of the gravitational
wave these two signal-to-noise ratios coincide.}:
\be
\rho\equiv2\sqrt{\int_0^{\infty} df \frac{|h(\theta_{\mathrm{true}})|^2}{S(f)}}\,.
\ee
The terms $\Delta \theta^a _{prior}$ and $\Delta \theta ^a _{est}$ are
the width of the prior distribution and of the Fisher matrix
1-$\sigma$ uncertainty estimate for the additional AG parameter,
respectively. Eq.~(\ref{odds approx}) given here is valid for the case
in which a zero realisation of the noise is present in the data. In Ref.~\cite{vallisneri:2012} noise is considered and
the appropriate formulae for the distribution of the Bayes factor over noise realisations can be found there. We opted for a zero-noise case for ease of comparison.

\section{Comparison between the exact calculation and the analytical
  approximation}\label{sec:results}

We compare the prediction from Eq.~(\ref{odds approx}) with the
evidence calculated by the Nested Sampling algorithm \cite{skilling:2004} as implemented in
\texttt{lalinference} \cite{lalinference} in a simple experiment. Using the
test waveform model presented in \cite{li:2012a}, we generate inspiral signals
which would span a range of FFs. The testing waveform is a frequency-domain stationary phase approximation waveform, based on the
TaylorF2 approximant \cite{buonanno:2009}, that has been modified in such way that the post-Newtonian
coefficients are allowed to vary around the GR values within a given
range. The TaylorF2 waveform for a face-on, overhead binary is given by:
\be
h(f) = \frac{1}{D}\sqrt{\frac{5}{24}}\pi^{-2/3}\mathcal{M}^{5/6}f^{-7/6}e^{i\Psi(f)},
\ee
where $D$ is the luminosity distance, $\mathcal{M}$ is the chirp mass
and the phase $\Psi(f)$ is:
\begin{eqnarray}
\Psi(f) &=& 2\pi f t_c - \phi_c - \pi/4 \\
&+& \sum_{i=0}^7 \left[\psi_i + \psi_{i}^{(l)} \ln f \right]\,f^{(i-5)/3}.\nonumber
\end{eqnarray}

The explicit forms of the coefficients $\psi_i$ and $\psi_{i}^{(l)}$
in $(\mathcal{M},\eta)$, where $\eta$ is the symmetric mass ratio, can be found in \cite{mais10}.
In all our experiments we kept the parameters of the simulated
sources fixed with the exception of the 1.5 post-Newtonian
coefficient $\psi_3$ which we varied between $[0.5,1.5]$ times its GR value
by adding an arbitrary shift $d\chi_3$ between $[-0.5,0.5]$:
\be
\psi_3 \rightarrow \psi_3(1+d\chi_3)\,.
\ee

\begin{figure}
\includegraphics[scale=0.5]{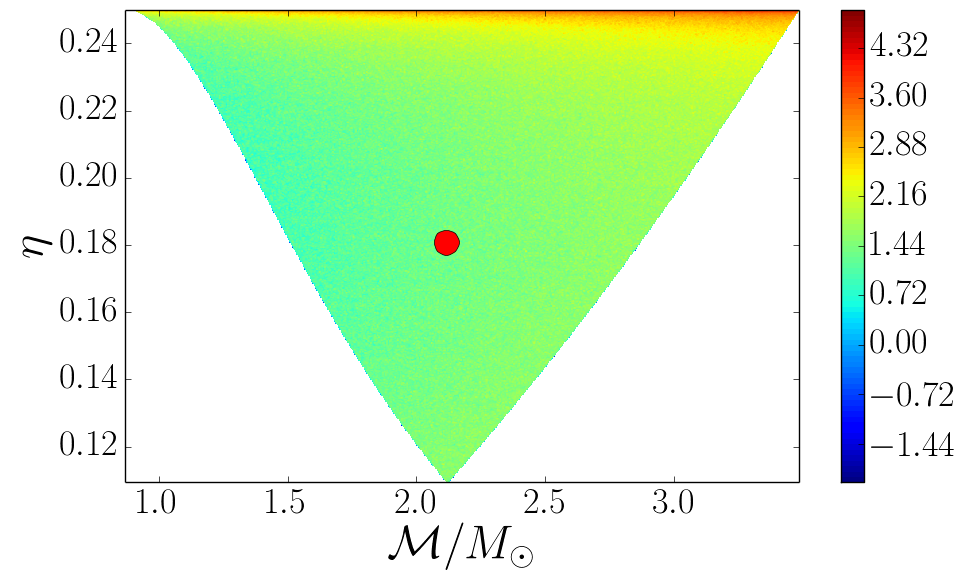}
\caption{Two-dimensional logarithmic prior probability distribution in the $\mathcal{M}$ and $\eta$
  space. The shape of the region is set by the allowed ranges of individual and total masses.  The
  red marker indicates the location of the $1.4M_\odot + 4.5M_\odot$ system
  simulated for the analysis.\label{fig:prior}}
\end{figure}

The Nested Sampling algorithm was set up to sample from the following
prior:
\begin{itemize}
\item the component masses where allowed to vary uniformly $\in
[1,7]M_\odot$ with the total mass constrained to the range $\in
[2,8]M_\odot$. This choice results in an allowed region of triangular shape in the
$\mathcal{M},\eta$ plane, see Fig.~\ref{fig:prior};
\item uniform on the 2-sphere for sky position and orientation
  parameters; 
\item uniform in Euclidean volume for the luminosity distance;
\item for recovery with AG templates, we used only one free testing parameter
  ($d\chi_3$) which was allowed to vary uniformly between $[-0.5,0.5]$
  times its GR value.
\end{itemize}

The FFs were computed from the maximum likelihood values obtained from
the \texttt{lalinference} simulations, see Appendix A. The parameter uncertainty for Eq.~(\ref{odds approx}) was computed
using a 5-dimensional Fisher matrix calculation in which we varied the two mass
parameters, the time of coalescence, the phase at coalescence and the deviation parameter $d\chi_3$. 

Our experiments were performed analysing simulated signals from a
system whose component masses were chosen to be $1.4M_\odot +
4.5M_\odot$. We chose this system because it lies in the centre
of our prior probability distribution over the masses, far away from prior boundaries.
This minimises the impact of the prior on the fitting factor and Bayes factor computations, which ensures that we can make a fair comparison with Eq.~(\ref{odds approx}) derived under the assumption of a uniform prior.%
\footnote{For example, an equal mass system would lie exactly on
the prior boundary at $\eta=0.25$. For this reason, the GR model has very little
room in the $\eta$ direction to accommodate the additional phase
shift due to a non-zero $d\chi_3$.  The net result is a very rapid drop in FF towards negative $d\chi_3$.}

The approximate formula in Eq.~(\ref{odds approx}) depends essentially
on two quantities: the signal-to-noise ratio $\rho$ and the FF.   Below, we describe our investigations of the dependence of the Bayes factor on these two quantities.

For Nested Sampling calculations, the uncertainty on the calculated value of the
evidence $Z$ is evaluated as \cite{skilling:2004}:
\begin{equation}\label{eq:dlogz}
\Delta\log Z \simeq \sqrt{\frac{H}{n}}
\end{equation}
where $H$ is the Kullback--Leibler divergence (or relative entropy)
between the posterior distribution and the prior distribution
and $n$ is the number of live points used for Nested Sampling.
$H$ is computed by the Nested Sampling along side the
evidence $Z$. Tipical values for $\Delta\log Z$ are  $O(10^{-1})$.

Finally, it was recently pointed out that when a signal terminates
abruptly in the detector band, measurement uncertainty may be significantly
smaller than predicted by the Fisher matrix calculation
\cite{mandeletal:2014}. 
To avoid these complications, we limited our analysis to frequencies
between $30$ and $512$ Hz.

\subsection{Scaling with the signal-to-noise ratio}

\begin{figure}
\includegraphics[scale=0.5]{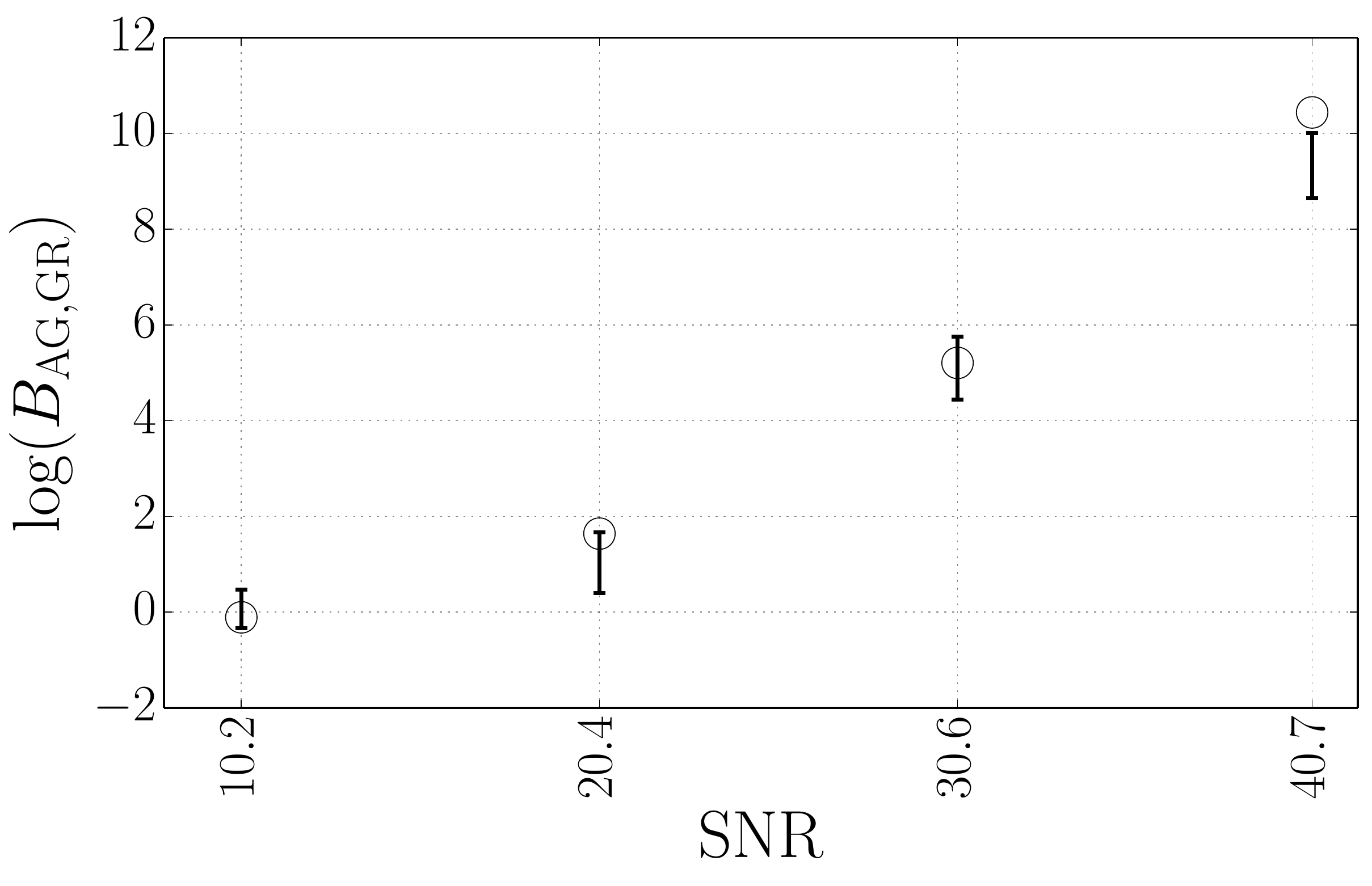}
\caption{Logarithmic Bayes factor from \texttt{lalinference} (errorbars)
  and from Eq.~(\ref{odds approx}) (empty circles) as a function of the SNR, 
   for FF fixed to 0.992. The \texttt{lalinference} errors are
   computed from Eq.~(\ref{eq:dlogz}). The circles are the values of the logarithmic Bayes factor obtained from Eq.~(\ref{odds approx}) using the FF extracted from the maximum likelihood values
  as computed by \texttt{lalinference}. \label{fig:snr}}
\end{figure}

We investigated the dependence of the Bayes factor on SNR by 
comparing the output of \texttt{lalinference} and Eq.~(\ref{odds
  approx}) for a $1.4 M_{\odot}+4.5M_{\odot}$ system at SNRs of 10, 20, 30 and
40 at a fixed value of the FF, $0.992$. Fig.~\ref{fig:snr} shows the
Bayes factors from the two calculations.

The quadratic dependence of the Bayes factor on the SNR was verified by
means of a simple chi-squared fit to an expression of the form 
\begin{equation}\label{eq:fit-snr}
\ln B_{\mathrm{AG},\mathrm{GR}} = \alpha \mathrm{SNR}^\beta +\gamma\,.
\end{equation} 
The scaling of $\log(B_{\mathrm{AG},\mathrm{GR}})$ with the SNR of the
 appears to be consistent with the expected quadratic dependence: we find $\beta=1.95\pm
0.4$.  


\subsection{Scaling with the fitting factor}

We evaluate the dependence of the Bayes factor on FF by again injecting a signal from a 
$1.4M_\odot + 4.5M_\odot$ binary, now at a fixed SNR of 20 but with varying FF. 
As in the previous section, we vary the FF by adding
arbitrary deviations from the GR value to the 1.5 post-Newtonian phase coefficient. In
particular, we varied $d\chi_3$ between $-0.5$ and $0.5$, leading to 
FF  $\in[0.7,1.0]$.  We verified that our injection was sufficiently far from prior boundaries by confirming that the Bayes factor is the same for positive and negative values of $d\chi_3$ that yield the same FF.

\begin{figure}
\includegraphics[scale=0.5]{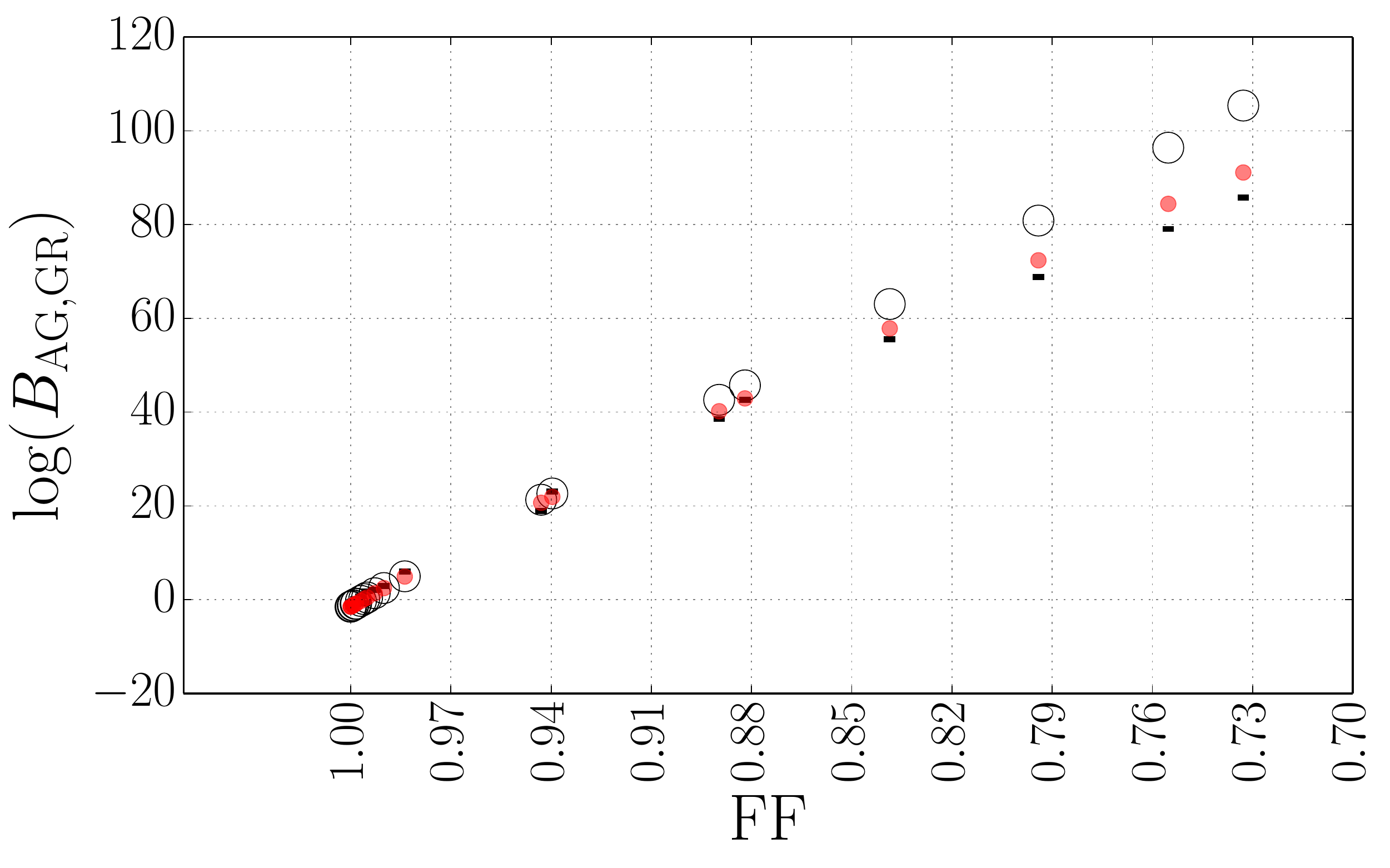}
\caption{Logarithmic Bayes factor from \texttt{lalinference} (error bars)
  and from Eq.~(\ref{odds approx}) (empty circles) as a function of the
  FF. The SNR was fixed to 20.   The Bayes factors computed with
  the two approaches agree for FF$\sim 1$ but tend to diverge for
  decreasing FF. 
  The red dots indicate the value of the logarithmic Bayes factor
  obtained by using a quadratic dependence on the FF, Eq.~(\ref{eq:odds-ff-2}), rather than the linear dependence of Eq.~(\ref{odds approx}).
  \label{fig:ff}}
\end{figure}

Fig.~\ref{fig:ff} shows the logarithmic Bayes factor computed by
\texttt{lalinference} and from Eq.~(\ref{odds approx}). The two
methods agree for FF $\sim 1$.  At FF $\leq 0.9$, the analytical
approximation overestimates the value of the Bayes factor compared to
\texttt{lalinference}. Moreover, the disagreement gets worse with
decreasing FF, suggesting a nonlinear dependence on
the FF.  In the next section, we investigate the approximate analytical expression in greater detail and derive additional corrections that extend its validity to lower fitting factors.  

\subsection{Correcting the analytical expression for lower fitting factors}

Under the assumption that the region of likelihood support on the
parameter space is small, and that over this region the prior
does not vary significantly, the evidence for any of the models $H_i$,
depending on parameters $\theta$, under
consideration can be approximated as (e.g., \cite{Veitch:2012}, correcting for a typo in the exponent of $(2\pi)$):
\be
Z(H_i) \propto[L_{H_i} ]_{\mathrm{max\,over\,} \theta} \, (2\pi)^{N/2} \, \prod_i^N \frac{\Delta \theta^i_{est}}{\Delta \theta^i_{prior}}\,,
\ee
where $N$ is the number of parameters.  Strictly speaking, the equation above is only valid when parameters are uncorrelated.  In the general case of correlated parameters, $\prod_i^N {\Delta \theta^i_{est}}$ should be replaced with the uncertainty volume in which the likelihood has support, while $\prod_i^N {\Delta \theta^i_{est}}$ is shorthand for the total prior volume.  However, such correlations do not affect the scaling of the uncertainty with the SNR, and do not impact our conclusions.

Therefore, the Bayes factor between the AG and GR model
can be approximated as the ratio of the maximum likelihoods times the
product of the ratios of posterior widths to prior supports:
\begin{equation}\label{Bayesfull}
B_{AG,GR} \approx \frac{[L_{AG}]_{\mathrm{max\,over\,}\theta^\prime}}{[L_{GR}]_{\mathrm{max\,over\,\theta}}} \sqrt{2\pi} \frac{\prod_{i=0}^N \frac{\Delta\theta^\prime_{est}}{\Delta\theta^\prime_{prior}}}{{\prod_{i=0}^{N-1}\frac{\Delta\theta_{est}}{\Delta\theta_{prior}}}}
\end{equation}
where $\theta^\prime$ and $\theta$ are parameter vectors within the AG and GR models, respectively, and $N$ is the dimensionality of the AG parameter space.

We begin by considering just the first term in Eq.~(\ref{Bayesfull}), which scales exponentially with the SNR in contrast to the components of the second term, which scale inversely with the SNR.  Neglecting the second term, we find:
\be\label{eq:deltalogl}
\log(B_{AG,GR}) \propto \log([L_{AG} ]_{\mathrm{max\,over\,} \theta^\prime})-\log( [L_{GR} ]_{\mathrm{max\,over\,} \theta})\,.
\ee
Using Eq.~(\ref{eq:ff-ns}), we find:
\be\label{eq:odds-ff-2}
\log(B_{AG,GR}) \propto \frac{\rho^2}{2} (1-FF^2)\,.
\ee
which is the expression originally proposed in \cite{cornish:2011}. At FF close to unity, $(1-\textrm{FF}^2) =(1+\textrm{FF})
(1-\textrm{FF}) \approx 2  (1-\textrm{FF})$, the approximation implicitly made in \cite{vallisneri:2012},
and we recover
Eq.~(\ref{odds approx}).  However, we expect (\ref{eq:odds-ff-2}) to
lead to a better fit at low fitting factors.  The filled (red) dots in Fig.~\ref{fig:ff} show the Bayes factor computed via Eq.~(\ref{eq:odds-ff-2}), with the proportionality constant fixed to be the same as in Eq.~(\ref{odds approx}). Indeed,
Eq.~(\ref{eq:odds-ff-2}) predicts Bayes factors that are in closer
agreement with the exact ones than Eq.~(\ref{odds
  approx}). In this case, disagreements with the exact result can be
seen for FF $\sim 0.75$, when the differences in the local shapes of the GR and AG manifolds can become significant. 

Vallisneri \cite{vallisneri:2012} further assumed that the priors and measurement uncertainties on all parameters except the one describing the deviation from GR, $\theta^a$, are the same for the AG and GR models (which, in turn, is a statement about the similarity in the shape of the two waveform manifolds near the maximum likelihood locations).  In this case, the Bayes factor between the two models is [cf.~(\ref{odds approx})]:
\be
B_{AG,GR} \propto \frac{[L_{AG} ]_{\mathrm{max\,over\,} \theta^\prime}}{ [L_{GR} ]_{\mathrm{max\,over\,} \theta}} \sqrt{2\pi} \frac{\Delta \theta^a_{est}}{\Delta \theta^a_{prior}}\,,
\ee
where $^a$ again refers to the one additional AG parameter which describes the deviation from GR.

However, we should not expect that the posterior widths will be identical in the AG and GR models for all parameters except the additional AG parameter are the same in the AG and GR models.  At high SNRs where the log likelihood can be approximated by a quadratic, posterior widths should scale inverse with the signal-to-noise ratio $\rho$.  While $\rho$ is the optimal SNR recovered when AG templates are used within the AG model, the maximal SNR recoverable when using GR templates within the GR model is lower.  By definition, this GR SNR is
\be
\rho_{\mathrm{GR}} \equiv  \left[ \frac{(h_{\mathrm{AG}}|h_{\mathrm{GR}}(\theta))}{\sqrt{ (h_{\mathrm{GR}}(\theta)|h_{\mathrm{GR}}(\theta))}}\right]_{\mathrm{max\,over\,} \theta} = \mathrm{FF}\, \rho\, .
\ee

Assuming the inverse SNR scaling of the posteriors, and using
identical priors on common parameters in the AG and GR models, Eq.~(\ref{Bayesfull}) reduces to
\be
B_{\mathrm{AG},\mathrm{GR}} \approx \frac{[L_{\mathrm{AG}}]_{\mathrm{max\,over\,}\theta^\prime}}{[L_{\mathrm{GR}}]_{\mathrm{max\,over\,\theta}}} \mathrm{FF}^{N-1}  \sqrt{2\pi} \frac{\Delta \theta^a_{est}}{\Delta \theta^a_{prior}}\,.
\ee
Taking a logarithm of this equation and again using $\frac{\rho^2}{2} (1-\mathrm{FF}^2)$ for the difference between maximum likelihoods (\ref{eq:ff-ns}), we find
\ba \label{eq:odds_prior_support}
\log(B_{\mathrm{AG},\mathrm{GR}}) &\approx& \frac{\rho^2}{2} (1-\mathrm{FF}^2) + (N-1)\log(\mathrm{FF}) \nonumber\\
&+& \log\left( 
\sqrt{2\pi} \frac{\Delta \theta^a_{est}}{\Delta \theta^a_{prior}} \right)\,.
\ea

\begin{figure}
\includegraphics[scale=0.5]{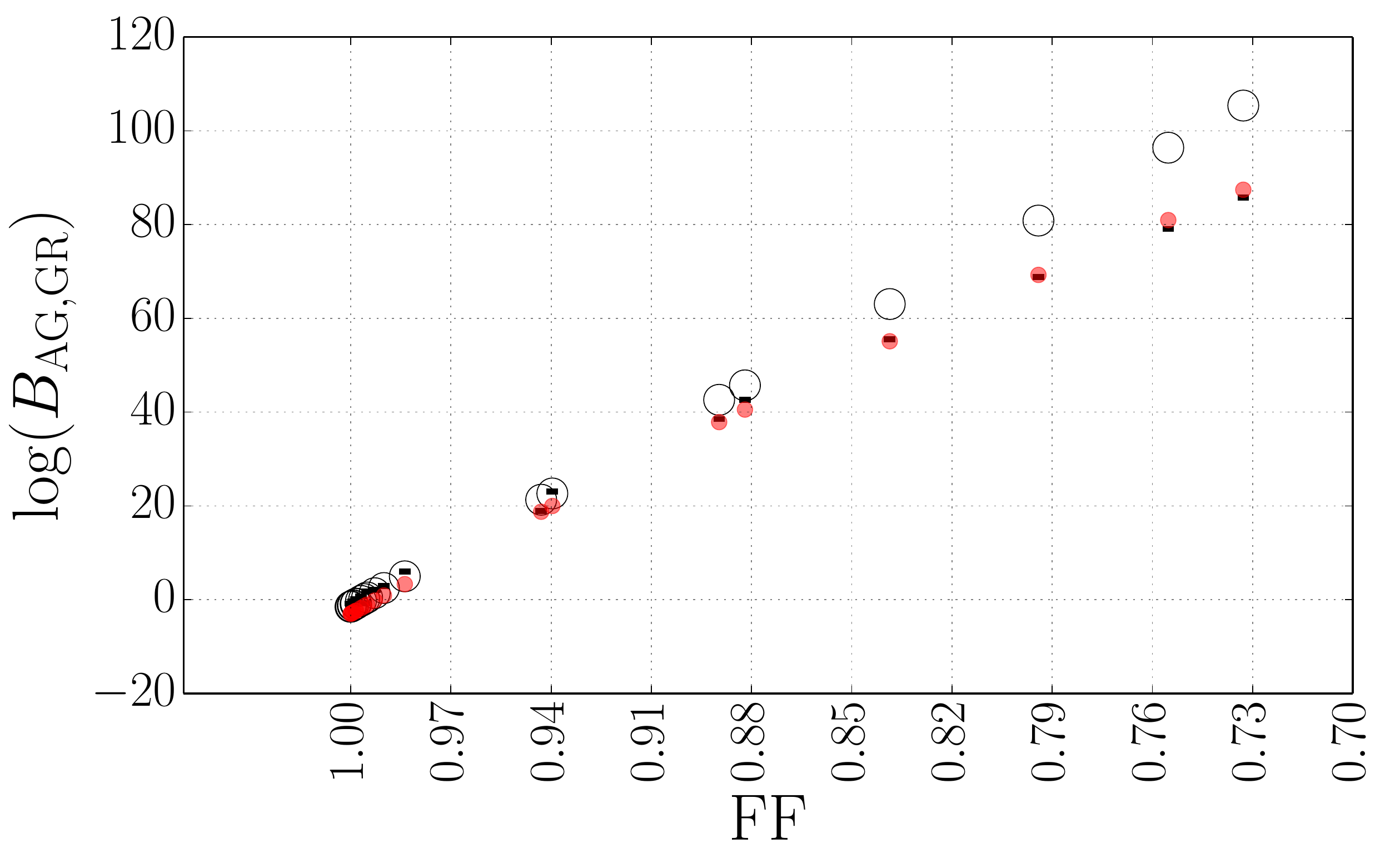}
\caption{Logarithmic Bayes factor from \texttt{lalinference} (error bars),
  from Eq.~(\ref{odds approx}) (empty circles), and from Eq.~(\ref{eq:odds_prior_support})
  (red dots) 
  as a function of the FF. The SNR is fixed to
  20. The \texttt{lalinference} errors are computed from
  Eq.~(\ref{eq:dlogz}).  The extended analytical expression of Eq.~(\ref{eq:odds_prior_support}) which includes corrections for lower fitting factors is in good agreement with the exact calculation from \texttt{lalinference}.\label{fig:prior_post}}
\end{figure} 

Eq.~(\ref{eq:odds_prior_support}) reduces to Eq.~(\ref{odds approx}) for $\mathrm{FF} \sim 1$.  However, it is accurate for a much wider range of fitting factors.  Fig.~\ref{fig:prior_post} shows the comparison between the
$\log$-Bayes factors from \texttt{lalinference} (error bars), the ones from
Eq.~(\ref{odds approx}) (circles) and finally the ones from
Eq.~(\ref{eq:odds_prior_support}) (red dots).  Indeed, the $\log$-Bayes
factors from Eq.~(\ref{eq:odds_prior_support}) show a very close
agreement with the numerical values.  Thus,
Eq.~(\ref{eq:odds_prior_support}) provides a good approximation to the
exact values of the $\log$-Bayes
factors. 

Another merit of Eq.~(\ref{eq:odds_prior_support}) is that it can be
generalised to an arbitrary number of extra non-GR parameters. If we
have $k$ non-GR parameters, Eq.~(\ref{eq:odds_prior_support}) becomes:
\ba 
\log(B_{\mathrm{AG},\mathrm{GR}}) &\approx& 
\frac{\rho^2}{2} (1-\mathrm{FF}^2)
+(N-k)\log(\mathrm{FF}) \nonumber \\
&+& \log\left( 
(2\pi)^{k/2} \prod_i^k\frac{\Delta \theta^{a_i}_{est}}{\Delta \theta^{a_i}_{prior}} \right)\,.
\label{eq:odds_multiparam}
\ea

Throughout this work, we have restricted our attention to the zero-noise realisation.  Vallisneri analysed the distribution of the Bayes factor under different noise realisations and showed  (see Eq.~(15) of \cite{vallisneri:2012}) that fluctuations in the logarithm of the Bayes factor have a standard deviation of $\sim \sqrt{2} \rho \sqrt{1-\mathrm{FF}}$.  While our additional corrections to the Bayes factor also lead to corrections in this quantity, we neglect these second-order effects.

We can compare the two systematic corrections discussed above to the level of these statistical fluctuations due to noise.  The difference in the log-Bayes factor between Eq.~(\ref{odds approx}) and Eq.~(\ref{eq:odds-ff-2}), i.e., the difference between the approximations of Refs.~\cite{vallisneri:2012} and \cite{cornish:2011}, is $(1/2) \rho^2 (1-\mathrm{FF}^2) - \rho^2 (1- \mathrm{FF}) = - (1/2) \rho^2 (1-\mathrm{FF})^2$.  This difference is approximately equal to the statistical fluctuation in the log-Bayes factor for $\rho=20$ and $\mathrm{FF} \sim 0.73$, corresponding to the rightmost points in Fig.~\ref{fig:prior_post}.  Meanwhile, the new correction to the log-Bayes factor which we introduced in Eq.~(\ref{eq:odds_prior_support}) has a magnitude of $(N-1) \log{\mathrm{FF}}$; for $N=9$ and other parameters as above, it is several times smaller than the noise-induced fluctuations.

Therefore, these corrections are unlikely to impact the detectability of a deviation from GR; in any case, in practice the detectability of the deviation would be determined by an analysis of the data and a numerical computation of the Bayes factor, not approximate predictive techniques.  However, these corrections are useful in explaining the apparent difference between numerical and analytical calculations, and therefore help validate both approaches by enabling a successful cross-check.

\section{Discussion}\label{sec:discussion}

We computed the Bayes factor between a GR model and an alternative
gravity model for a gravitational-wave signature of an inspiraling
compact binary.  We compared two calculations of the Bayes factor: an
exact numerical computation with \texttt{lalinference} and an
approximate analytical prediction due to Vallisneri
\cite{vallisneri:2012}.  We verified that the analytical approximation
yields the correct scaling of the logarithm of the Bayes factor with
the square of the signal-to-noise ratio at high fitting factor values.
However, the predicted scaling of the Bayes factor with the fitting
factor is inaccurate for FF $\leq 0.9$.

We extended the regime of validity of the analytical approximation of
\cite{vallisneri:2012} to lower fitting factors by including
additional FF-dependent terms and by extending to multiple non-GR
parameters.  We confirmed that the more complete analytical prediction
that we derived in this work, Eq.~(\ref{eq:odds_prior_support}),
remains valid down to fitting factors of $\leq 0.7$. 

It is worth noting that  Eq.~(\ref{odds approx}) loses accuracy precisely in the
regime where it becomes possible to differentiate GR and alternative gravity models. The FF is very close to unity in the regime in which the GR waveform can still match a signal which violates GR through different choices of the values of the binary's parameters within the GR
framework.   The Bayes factor in this case is not significantly different
from 1, thus no decision on the nature of the signal can be made at an
acceptably low false alarm probability\footnote{This regime is a case of the so-called
  ``fundamental bias'' \cite{yunespretorius09}. It is treated using
  the analytical approximation presented in \cite{vallisneri:2012} by
  \cite{vallisneriyunes13}. A numerical study with the 
  \texttt{lalinference} code can be found in
  \cite{vitale-delpozzo:2014}.}.  Therefore, our extension of the
analytical expression for the Bayes factor to lower fitting factors
provides a useful, easy-to-compute approximate technique precisely in
the regime of interest in the case of a zero noise realisation.
  
The analytical expressions presented in \cite{cornish:2011,vallisneri:2012} and in this work are predicated on the assumption that the $(N-1)$--dimensional GR
manifold and the $N$--dimensional AG manifold are sufficiently similar near the maximum-likelihood values that the parameter uncertainties can be assumed to be equal (up to scaling with the inverse SNR) on the two manifolds.  Differences in the local curvature of the two manifolds could become significant when the distance between them is large, or the systematic bias between true and best-fit parameters is significant relative to statistical measurement uncertainty.  Therefore, this assumption could (although need not) break down either at small fitting factors, or, somewhat paradoxically, at large SNR for a fixed fitting factor.  Specifically, when $\rho^2 (1-FF^2) \gg N$, the uncertainty region within a manifold is much smaller than the distance between manifolds or between true and best-fit parameters within a manifold, and the AG and GR manifolds may no longer yield similar parameter uncertainties.

Another possible cause of the breakdown of the analytical approximation is the impact of priors.  If the prior distribution is very non-uniform within the region of likelihood support, particularly if a sharp prior boundary is present within this region, the analytical approach described above is no longer valid.  A further limitation is the restriction to high SNRs.  The widths of the posterior
distributions are inversely proportional to the SNR only when the linearized-signal approximation is valid (i.e., when the covariance matrix is well approximated by the inverse of the Fisher matrix).

In summary, the analytical approximation presented by Cornish et al.~\cite{cornish:2011} and Vallisneri \cite{vallisneri:2012}, and its extensions as given in Eqs.~(\ref{eq:odds_prior_support}) and (\ref{eq:odds_multiparam}), provide a computationally cheap way of predicting the detectability of a deviation
from GR for a given AG theory without the need to run expensive
numerical simulations, subject to the limitations outlined above.  
 Hence, these analytical approximations can be a
very useful tool to get quick indications of whether a particular
class and magnitude of one-parameter deviations from GR are
detectable.  However, these methods are merely predictive, and inference on actual data must rely on parameter estimation and model comparison with complete data-analysis pipelines.

\section*{Acknowledgements}  
We thank Michele Vallisneri, Neil Cornish, John Veitch, Will Farr, Christopher Berry, Carl-Johan Haster and
Zachary Hafen for useful comments and discussions.
The work was funded in part by a Leverhulme Trust research project grant.
The numerical simulations were performed on the Tsunami cluster of the
University of Birmingham.\\

\appendix
\section{Computing Fitting Factors from logLikelihoods}
The fitting factor Eq.~(\ref{eq:ff}) can be extracted directly from
the Nested Sampling runs without the need to search over a parameter grid. 
Begin by writing the logarithmic likelihood in a zero noise realisation:
\be
\log(L) = const+(h_{true}|h(\theta))-\frac{(h_{true}|h_{true})}{2}-\frac{(h(\theta)|h(\theta))}{2}
\ee 
where $h_{true}$ is the gravitational wave signal in the data stream, $h(\theta)$ is the search template,  and $const$ is a constant.  Consider the difference $\Delta\lambda$ between the maximum log likelihoods given for the AG and GR models given an AG signal:
\begin{equation}
\Delta\lambda = \log(L_{GR})_{\mathrm{max\,over\,}
\theta}-\log(L_{AG})_{\mathrm{max\,over\,} \theta^\prime}\,,
\end{equation}
where
\begin{eqnarray}
\log(L_{GR}) &=&
const+(h_{true}|h_{GR}(\theta)) \label{LGR}\\
&&-\frac{(h_{true}|h_{true})}{2}\nonumber\\
&&-\frac{(h_{GR}(\theta)|h_{GR}(\theta))}{2} \nonumber
\end{eqnarray}
and $\log(L_{AG})_{\mathrm{max\,over\,} \theta^\prime} = const$ since this likelihood is maximized for
$h_{AG}(\theta^\prime) = h_{true}$.
 
We can maximise the GR log-likelihood analytically over the amplitude
of $h_{GR}(\theta)$ by defining
\be
h_{GR}(\theta) = A \hat{h}_{GR}(\xi)
\ee
with $\xi \equiv \theta \setminus A$ being the set of parameters other than the amplitude. One can solve for the value of the amplitude that satisfies
\be
\frac{\partial \log(L_{AG})}{\partial A}=0
\ee
and obtain:
\be
A=\frac{(h_{true}|\hat{h}_{GR}(\xi))}{(\hat{h}_{GR}(\xi)|\hat{h}_{GR}(\xi))}\,.
\ee
Substituting this into Eq.~(\ref{LGR}) and setting $(h_{true}|h_{true}) \equiv \rho^2$ yields:
\be
\Delta\lambda =\frac{1}{2}[\frac{(h_{true}|\hat{h}_{GR}(\xi))^2}{(\hat{h}_{GR}(\xi)|\hat{h}_{GR}(\xi))}]_{\mathrm{max\,over\,}
\xi}- \frac{\rho^2}{2}\,.
\ee 

Meanwhile, the fitting factor FF can be similarly written as:
\be
FF=[\frac{(h_{true}|\hat{h}_{GR}(\xi))}{\rho\sqrt{(\hat{h}_{GR}(\xi)|\hat{h}_{GR}(\xi))}}]_{\mathrm{max\,over\,}
\xi} \,.
\ee
Thus, 
\be \label{Deltalambda}\label{eq:ff-ns}
\Delta\lambda  = \frac{1}{2}\rho^2 FF^2 - \frac{\rho^2}{2} = - \frac{\rho^2}{2} (1-FF^2)
\ee
and
\be
FF = \sqrt{\frac{2\Delta\lambda}{\rho^2}+1}\,.
\ee

The numerically computed values of $\Delta\lambda$ have an intrinsic variability due to the
stochastic nature of the sampler. The standard deviation for
$\Delta\lambda$ derived from our simulations is $\sigma_{\Delta \lambda} = 0.016$. The
corresponding uncertainty in our FF estimate is given by 
\be
\sigma_{FF} = \frac{1}{\rho^2 FF}\sigma_{\Delta \lambda}\,.
\ee
For an SNR of 20 and a FF of 1, $\sigma_{FF}=4\times 10^{-5}$. For all
practical purposes we can consider our estimated FFs to be exact. 
\end{document}